\documentclass[%
  reprint,
  aps,
  pra,
  longbibliography,
  superscriptaddress,
  nofootinbib
]{revtex4-2}

\usepackage{amsmath,amssymb,mathtools,bm}
\usepackage{physics}
\usepackage{siunitx}
\usepackage{microtype}
\usepackage{graphicx}
\usepackage[colorlinks=true,allcolors=blue!60!black]{hyperref}
\usepackage{dblfloatfix}
\usepackage{balance}
\usepackage{xcolor}
\usepackage{algorithmic}
\usepackage{graphicx}
\providecommand{\myfigscale}{0.75} 

\newcommand{\kk}{\mathbf{k}_\perp}
\newcommand{\rr}{\mathbf{r}_\perp}
\newcommand{\XX}{\Delta\mathbf{X}}

\usepackage{tikz}
\usetikzlibrary{arrows.meta,positioning,fit,calc,shapes.geometric}
\tikzset{
  >={Latex[length=3mm]},
  block/.style={draw, rounded corners=2pt, minimum height=6mm, inner xsep=4mm, font=\small},
  light/.style={line width=0.6pt},
  note/.style={font=\scriptsize, align=center}
}

\begin{document}

\title{FRINGE: a protocol for self-referenced quantum state estimation via photon-number-resolved interferometry}

\author{Matan Even Tzur}
\affiliation{Max Planck Institute for the Structure and Dynamics of Matter, Luruper Chaussee 149, 22761 Hamburg, Germany}


\date{\today}

\begin{abstract}
We introduce a \emph{self-referenced} method for quantum–state tomography of light based on photon-number–resolved double-slit interferometry. Two identical copies of the unknown quantum field illuminate laterally displaced slits, guaranteeing perfect spatiotemporal mode matching \emph{without} a separate local oscillator. In the far-field, detection at transverse position $x$ is associated with a relative slit phase $\phi(x)$, and an $N$-photon event projects the detected quantum field onto a state $\ket{N;\phi(x)}$. The resulting  distribution $P(N,\phi)$ is the quantum analogue of a Frequency Resolved Optical Gating (FROG) trace: whereas FROG reconstructs the classical complex spectral field $E(\omega)$ from a spectrally resolved second harmonic of a pulse with its delayed self, our measurement reconstructs the Fock-space wavefunction or density matrix from binomially weighted self-interference. The scheme requires no known or mode-matched reference and is compatible with commercially available photon-number-resolving cameras. Beyond conceptual simplicity and automatic mode matching, the FROG analogy permits direct transfer of mature ultrafast-optics methodologies--e.g., mixed-state, ptychographic, and vectorial extensions--into quantum optics, offering a versatile route to tomography of quantum photon states.
\end{abstract}

\maketitle
\textbf{FRINGE -- Fock-Resolved Interferometry for Number-Gated quantum state Estimation.}
Quantum-state tomography of optical fields most commonly uses balanced homodyne detection: a signal is interfered with a strong, phase-stable local oscillator (LO), the quadrature statistics are sampled versus LO phase, and the state is then reconstructed numerically--e.g., via inverse Radon transforms of the measured distributions--to obtain the Wigner function or density matrix \cite{LvovskyRaymerRMP2009,Leonhardt1997,SmitheyPRL1993}. Homodyne detection has enabled reconstructions of single-photon states \cite{Fock_LvovskyPRL2001,DFock_LvovskyBabichevPRA2002}, displaced Fock states \cite{DFock_LvovskyBabichevPRA2002}, and squeezed states \cite{Squeezed_BreitenbachNature1997} even under imperfect detection efficiency \cite{LowEff_EspositoNJP2014}. More recently, \emph{electro-optic sampling} (EOS) has emerged as a time-domain, field-sensitive alternative in which an ultrashort probe pulse samples sub-cycle field fluctuations via the Pockels effect; by scanning delay and analyzing probe polarization, one can reconstruct quadrature statistics without a conventional LO \cite{RiekScience2015,riek2017subcycle,hubenschmid2024optical}. A practical requirement shared by many such quantum state tomography schemes is a \emph{mode-matched reference} (e.g., an LO or an ultrashort probe) whose spatiotemporal mode structure closely matches that of the quantum signal. In many quantum light sources, however, the nonlinear generation process significantly distorts the signal’s spatial and temporal modes, often in ways that are difficult to control, making high-fidelity mode matching technically demanding and frequently a dominant limitation to reconstruction accuracy.

A complementary route is self-referenced tomography: interfere (or gate) a state
with an identical copy of itself, produced by duplicating the optical path, so that both copies share the same spatiotemporal mode by construction. Self-referencing has a long pedigree in ultrafast optics, beginning with intensity autocorrelation for coarse pulse-length estimates, which later led to frequency-resolved optical gating (FROG) for full field retrieval \cite{TrebinoBook2002,TrebinoJOSAA1993}. In a typical second-harmonic FROG experiment, an ultrashort pulse is combined with a delayed copy of itself in a $\chi^{(2)}$ crystal, and one measures the spectrum of the generated second harmonic versus delay,
\begin{equation}
S(\omega,\tau)=\Big|\int dt\,E(t)\,E(t-\tau)\,e^{-i\omega t}\Big|^2
\label{eq:frog}
\end{equation}
with $E(t)$ the electric field, $\tau$ the delay, and $\omega$ the angular frequency\cite{TrebinoBook2002}. Unlike autocorrelation, the added spectral dimension renders the trace informationally complete: it encodes sufficient redundancy to recover both amplitude and phase of $E(t)$ across its spectral support. Rigorous analyses show that the FROG trace uniquely determines the pulse up to the well-known  ambiguities: global phase, absolute time shift, and time reversal \cite{TrebinoBook2002,bendory2017uniqueness}. 

\begin{figure}[t]
  \centering
  \includegraphics[width=\myfigscale\linewidth]{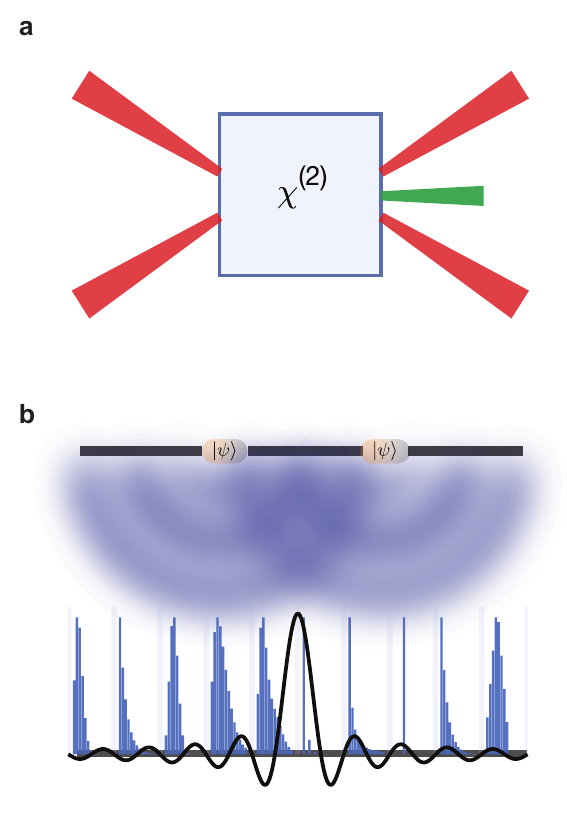}
  \caption{Schematic illustration of \textbf{(a)} Frequency-Resolved Optical Gating (FROG) for reconstructing ultrashort laser pulses and \textbf{(b)} Fock-Resolved Interferometry for Number-Gated quantum-state estimation (FRINGE). In FROG, $E(t)$ and $E(t-\tau)$ are mixed in a nonlinear crystal, and the resulting spectrogram enables reconstruction of $E(t)$. In FRINGE, a quantum state interferes with a displaced (or delayed) copy of itself and the photon-number distribution is measured in the far field. The two experiments are mathematically analogous, allowing decades of advances from ultrafast optics to be translated to quantum optics.}
  \label{fig:FROGvsFRINGE}
\end{figure}

Here we  show that a photon–number–resolved (PNR) double-slit experiment forms a quantum-optical analogue of FROG and therefore inherits its chief advantages (self-referencing, informational completeness), as well as the same ambiguities~\cite{TrebinoBook2002}. Whereas FROG reconstructs the complex spectral field \(E(\omega)\), the PNR double-slit measurement reconstructs the Fock-space probability amplitudes \(\{c_n\}\) (the probability amplitudes of the \(n\)-photon components) or, more generally, the density matrix \(\rho_{nm}\) (Fig.~\ref{fig:FROGvsFRINGE}).  We refer to this approach as FRINGE - Fock-Resolved Interferometry for Number-Gated quantum-state Estimation.

A concrete motivation for FRINGE comes from recent experimental demonstrations of photon–number–resolved interferometry with identical replica beams in the extreme ultraviolet spectral range (XUV), where the interfering fields are high–harmonic generation (HHG) beams~\cite{harrison2022increased,harrison2023generation,trallero2025squeezed}. Despite decades of work, the quantum state of HHG light has never been reconstructed: no practical local oscillator (LO) exists at XUV wavelengths, and the only fields that can be mode–matched to an HHG beam are other HHG beams whose quantum state is itself unknown. These attosecond–interferometry experiments already realize precisely the ingredients required for a self–referenced protocol: the generation of identical XUV HHG beams in a Young’s double–slit geometry~\cite{harrison2022increased,harrison2023generation} and photon counting of the resulting XUV interference pattern~\cite{trallero2025squeezed}. That is, the configuration analyzed is already within experimental reach, providing a natural platform on which to implement FRINGE and, for the first time, perform quantum–state tomography of XUV high harmonics. More broadly, such self–referencing offers an initial strategy for quantum state estimation in any spectral regime that lacks a quantum–characterized LO: once a state is validated as approximately coherent via FRINGE, it can itself serve as the LO for conventional homodyne quantum–state tomography.

We begin by considering the geometry presented in Figure~\ref{fig:FROGvsFRINGE}(b). Two slits (indexed by \(j=1,2\)) are centered at transverse positions $x=\pm\Delta X/2$ in the object plane $z=0$ and separated along the $x$ axis by a distance $\Delta X$.  A photon-number-resolving camera is placed at distance $Z$ in the far field.  We denote by $\rr=(x,y)$ the transverse coordinate in the slit plane, and by $x$ the transverse coordinate along the slit separation. 
This geometry is illuminated with two quasi–paraxial fields centered at an angular frequency \(\omega_0\), occupying a product state \(\ket{\psi}_1\otimes\ket{\psi}_2\) with \(\ket{\psi}_j=\sum_{n\ge0}c_n\ket{n}_j\), where \(\ket{n}_j\) correspond to an \(n\) photon state of the \(j=1,2\) slit. Formally, \(|n\rangle_j\) are Fock states of the annihilation operators
\begin{equation}
\hat a_j = \int d\omega \int d^2\kk\,
\tilde u_j^*(\kk,\omega)\,\hat a(\kk,\omega),
\qquad j=1,2 .
\label{eq:mode-expansion}
\end{equation}
where \(\tilde u_j(\kk,\omega)\) is the (normalized) spatiotemporal mode function of slit \(j\),
\begin{equation} 
\int d\omega \int d^2\mathbf{k_\perp} \big|\tilde u_j(\mathbf{k_\perp},\omega)\big|^2 = 1 , \label{eq:mode-normalization} \end{equation}
and the plane-wave operator densities \(\hat a(\kk,\omega)\) obey
\begin{equation} 
\big[\hat a(\mathbf{k_\perp},\omega),\hat a^\dagger(\mathbf{k'_\perp},\omega')\big] = \delta(\omega-\omega')\delta^{(2)}(\mathbf{k_\perp}-\mathbf{k_\perp'}) . \label{eq:comm-rel} 
\end{equation}
Assuming identical spatiotemporal profiles, the two slit modes differ only by the lateral shift $\Delta X$ along $x$ in the object plane, so that the spatial mode functions obey $u_2(\rr,\omega)=u_1(\rr-\XX,\omega)$.
The mode commutator is the overlap integral,
\begin{equation} 
\big[\hat a_i, \hat a_j^\dagger\big] =
\int d\omega \int d^2\rr\, u_i^*(\rr,\omega)\,u_j(\rr,\omega) .
\label{eq:comm-modes} 
\end{equation}
and for well–separated identical slits \( [\,\hat a_i,\hat a_j^\dagger\,]\approx\delta_{ij}\). Under paraxial (Fraunhofer) propagation, the far field is the Fourier transform of the object (slit) plane; the lateral shift by \(\Delta X\) multiplies the angular spectrum by a linear phase, $\tilde u_2(q_x)=e^{\,i q_x\Delta X}\tilde u_1(q_x)$, where $q_x$ is the transverse wave–vector component conjugate to $x$ (with $q_x\simeq k_0 x/Z$ in the paraxial approximation). This yields the position–phase relation  \cite{GoodmanFO3e}
\begin{equation}
  \phi(x)\;\simeq\;\frac{k_0}{Z}\,x\,\Delta X,\qquad k_0=\frac{\omega_0}{c}.
\end{equation}
A PNR click of total photon number \(N\) at screen position \(x\) projects onto the detection mode
\begin{subequations}\label{eq:PNR-projection}
\begin{equation}
  \ket{N;\phi(x)}\equiv\frac{1}{\sqrt{N!}}\Big[\hat c^\dagger(\phi(x))\Big]^N\ket{0},
\end{equation}
\begin{equation}
  \hat c^\dagger(\phi)\equiv\frac{\tilde u_1(q_x)}{\sqrt{2}}\Big(\hat a_1^\dagger+e^{-i\phi}\hat a_2^\dagger\Big).
\end{equation}
\end{subequations}
For simplicity, we approximate the angular spectrum as flat over the detected region, $\tilde u_1(q_x)\approx 1$; the impact of imperfect mode matching is discussed in Sec.~I of the SI. Then, the probabilities to measure $N$ photons at the detection position $x$ (phase $\phi(x)$) is: 

\begin{equation}
  P(N,\phi)\;=\;\big|\frac{1}{2^{N/2}}\sum_{m=0}^{N}\sqrt{\binom{N}{m}}\;e^{\,i m \phi}\;c_{N-m}\,c_m\big|^2.
  \label{eq:P(N,phi)}
\end{equation}

The function \(P(N,\phi)\) exhibits the same mathematical structure of Equation \eqref{eq:frog}. The role of time $t$ is played by the single-slit Fock index $m$, the delay $\tau$ by the total photon number $N$, and the optical frequency $\omega$ by the interferometric phase $\phi$. The dictionary is:

\begin{center}
\renewcommand{\arraystretch}{1.2}
\begin{tabular}{c|c}
\textbf{FROG (SHG)} & \textbf{Two-slit, PNR interferometry}\\\hline
$E(t)$ & $\{c_m\}_{m\ge 0}$ \\
$E(t-\tau)$ & $\{c_{N-m}\}_{m=0}^N$ \\
$t$ & $m$ \\
$\tau$ & $N$ \\
$\omega$ & $\phi$ \\
$E(t)E(t-\tau)$ & $\sqrt{\binom{N}{m}}\;c_{N-m}c_m$ \\
$\mathcal F_t[\cdot](\omega)$ & $\sum_m (\cdot)\,e^{i m\phi}$ \\
$|\cdot|^2$ & $|\cdot|^2$
\end{tabular}
\end{center}

As in FROG, redundancy across two axes ($N$ and $\phi$) compensates for the lost phase and renders the measurement informationally complete up to trivial ambiguities. The ambiguities in $P(N,\phi)$ are:
\begin{subequations}\label{eq:ambigs}
\begin{align}
c_n &\mapsto e^{i(\chi+n\theta)}\,c_n, \label{eq:ambigs-ramp}\\
c_n &\mapsto c_n^\ast. \label{eq:ambigs-conj}
\end{align}
\end{subequations}
These ambiguities have a simple interpretation in terms of the Wigner function $W(X,P)$, the phase-space quasiprobability distribution of the state, where $X$ and $P$ are the dimensionless field quadratures (real and imaginary parts of the complex amplitude). The global phase $\chi$ is unobservable. The linear phase ramp $e^{in\theta}$ rotates the Wigner function about the phase-space origin by angle $\theta$, corresponding to a time shift of the field which cannot be determined without an external reference. The conjugation $c_n \mapsto c_n^*$ reflects the Wigner function across the $X$-quadrature-axis (i.e., $P \mapsto -P$).

\begin{figure}[t]
  \centering
  \includegraphics[width=\linewidth]{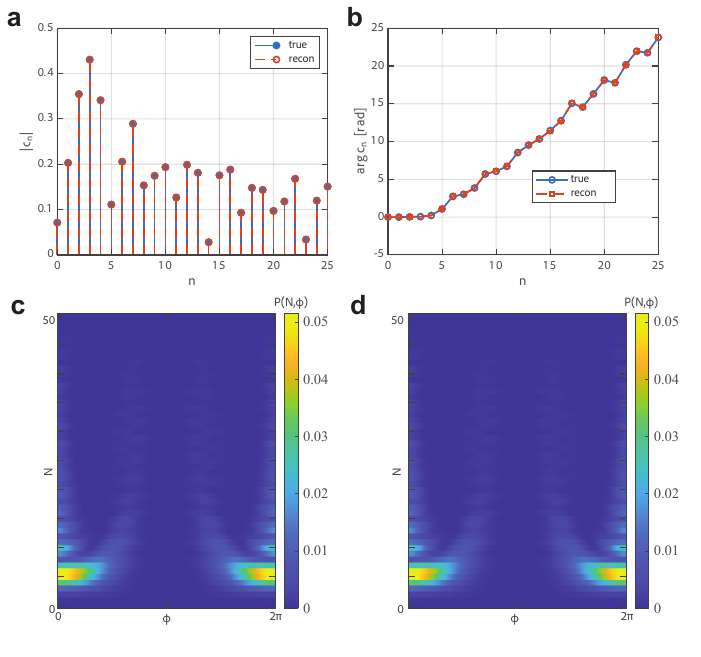}
 \caption{Reconstruction of single-slit Fock coefficients from interferometric number–gated data.
\textbf{(a)} Magnitudes $|c_n|$ and \textbf{(b)} phases $\arg c_n$ of a single-slit  squeezed state $D(\alpha)S(\zeta)\ket{0}$, recovered from the true distribution $P(N,\phi)$.
Blue circles: ground truth $c_n$ coefficients of the squeezed–coherent state.
Orange rectangle: FRINGE-reconstructed $c_n$ coefficients; the gauge is fixed by $c_0>0$ and $\arg c_1=0$. \textbf{(c)} Synthetic FRINGE data generated from the ground truth squeezed–coherent single-slit state.
\textbf{(d)} \(P(N,\phi)\) prediction from the \emph{reconstructed} single-slit coefficients $\{c_n\}$ obtained by our FRINGE procedure.
The overlap across $n$ demonstrates accurate recovery of both amplitudes and relative phases.}
  \label{fig:true-vs-recon}
\end{figure}

To recover \({c_n}\), we Fourier-analyze the fringe pattern at each photon number \(N\). Since $\phi(x) \propto x$, the Fourier transform along the screen coordinate $x$ is equivalently an integral over $\phi$:
\begin{equation}
\tilde P(N,\ell)
:=\frac{1}{2\pi}\!\int_{0}^{2\pi}\!P(N,\phi)\,e^{-i\ell\phi}\,d\phi
\quad(\ell\in\mathbb{Z}).
\label{eq:Phat-def}
\end{equation}
By Equation \eqref{eq:P(N,phi)}, 
\begin{equation}
\tilde P(N,\ell)
=2^{-N}\!\!\sum_{\substack{0\le m,m'\le N\\ m-m'=\ell}}
\sqrt{\binom{N}{m}\binom{N}{m'}}\;
c_{N-m}\,c_m\,c_{N-m'}^{*}\,c_{m'}^{*}.
\label{eq:Phat-exp}
\end{equation}
From this formula, we can directly obtain the magnitudes $|c_n|$:
\begin{subequations}\label{eq:c-mag}
\begin{align}
|c_0| &= \tilde P(0,0)^{1/4}, \label{eq:c0-mag}\\[2pt]
|c_1| &= \frac{\sqrt{|\tilde P(1,0)|}}{|c_0|}, \label{eq:c1-mag}\\[2pt]
|c_N| &= \frac{\sqrt{\,2^{N}\,|\tilde P(N,N)|\,}}{|c_0|}, \qquad N\ge 2. \label{eq:cN-mag}
\end{align}
\end{subequations}
To recover the relative phases, write $c_N=|c_N|e^{i\phi_N}$ and fix $\phi_0=\phi_1=0$ -- essentially setting the unobservable global phase $\chi$ and linear phase ramp $\theta$ in Equation \eqref{eq:ambigs} and setting orientation of the Wigner function of the quantum state in optical phase space.
With this convention, the nearest–neighbor phase increment $\Delta_N:=\phi_N-\phi_{N-1}$ is obtained, for $N\ge2$, from the $(N\!-\!1)$ harmonic as
\begin{equation}
\Delta_N \;=\; \arccos\!\left(
\frac{\tilde P(N,N-1)}{2^{\,1-N}\sqrt{N}\,|c_0 c_1 c_{N-1} c_N|}
\right),
\qquad N\ge2.
\label{eq:DeltaN}
\end{equation}
Because $P(N,\phi)$ is invariant under global conjugation $c_N\!\mapsto\!c^*_N$, the initial increment $\Delta_2$ is undetermined up to a sign (Equation \eqref{eq:DeltaN} has a sign ambiguity for $\Delta_N$). To resolve it, we arbitrarily choose $\Delta_2>0$, which in turn determines the sign of $\Delta_N$ for $N\ge3$. The true sign of $\Delta_{N}$ is the one that satisfies an additional relationship derived from Equation \eqref{eq:Phat-exp}:
\begin{equation}
\cos\!\big(\Delta_{N-1}+\Delta_N-\Delta_2\big)
=\frac{2^{N-1}\,\tilde P(N,N\!-\!2)-\tfrac{N}{2}\,|c_1|^2|c_{N-1}|^2}
{\sqrt{\binom{N}{2}}\,|c_0|\,|c_2|\,|c_{N-2}|\,|c_N|}.
\label{eq:cos-endpoint-scaled-final}
\end{equation}
To numerically validate our scheme we consider the single–slit state to be the squeezed–coherent state \(\ket{\psi_{\rm sc}}=D(\alpha_{\rm sc})S(\zeta_{\rm sc})\ket{0}\) where $D(\alpha_{sc})$ and $S(\zeta_{sc})$ are the standard single‑mode displacement and squeezing operators, and \(\alpha_{\rm sc}=\sqrt5e^i\), \(\zeta_{\rm sc}=1.5e^{0.3i}\). The magnitudes and phases of the corresponding reference Fock coefficients used to generate the synthetic data (the "ground truth") Fock coefficients \(c_n=\langle n|\psi_{\rm sc}\rangle\) are shown in Figure \ref{fig:true-vs-recon} (blue). From these coefficients we synthesize the interferometric number–phase map \(P(N,\phi)\) via Equation \eqref{eq:P(N,phi)}, displayed in Figure \ref{fig:true-vs-recon}(c). Applying the closed-form inversion formulas of Equations \eqref{eq:c-mag}-\eqref{eq:cos-endpoint-scaled-final} to the same dataset yields the reconstructed coefficients \(c_n\). The reconstruction coincides with the ground truth in Figure \ref{fig:true-vs-recon} (orange), and a predicted \(P(N,\phi)\) trace based on the reconstructed state reproduces the true \(P(N,\phi)\) in Figure \ref{fig:true-vs-recon}(d), demonstrating accurate recovery of both \(|c_N|\) and the relative phases.

\begin{figure}[t]
  \centering
  \includegraphics[width=0.9\linewidth]{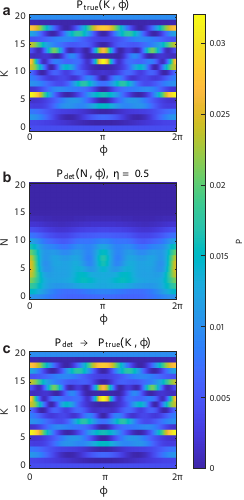}
\caption{
Interferometric number-phase maps $P(N,\phi)$ in the Young double-slit geometry (Figure \ref{fig:FROGvsFRINGE}b). \textbf{(a)} Ideal trace $P_{\rm true}(K,\phi)$ synthesized from a random quantum state.  \textbf{(b)} Detected trace $P_{\det}(N,\phi)$ after applying a finite quantum efficiency of $\eta=0.5$, which binomially thins the photon number statistics. \textbf{(c)} Exact reconstruction of the lossless trace $P_{\rm true}(K,\phi)$ obtained from panel~(b) by applying the closed-form inverse \eqref{eq:coef-inverse-final}.}
\label{fig:fringe-loss-inverse}
  \label{fig:QuantumEfficiency}
\end{figure}

\textbf{Detection efficiency. }
When the detectors have quantum efficiency \(\eta<1\), we do not measure the ideal lossless FRINGE trace \(P_{\rm true}(K,\phi)\) that would be obtained at unit efficiency. Instead, we detect the efficiency-degraded distribution \(P_{\det}(N,\phi)\), related to the ideal distribution \(P_{\rm true}(K,\phi)\) by  \cite{KissHerzogLeonhardtPRA1995}
\begin{equation}
P_{\det}(N,\phi)
=\sum_{K=N}^{N_{\max}} \binom{K}{N}\,\eta^N (1-\eta)^{K-N}\,
P_{\rm true}(K,\phi).
\label{eq:Pdet-forward-final}
\end{equation}
This relationship may be inverted to express $P_{\rm true}$ in terms of $P_{det}$ \cite{LeePRA1993,KissHerzogLeonhardtPRA1995,HerzogPRA1996,GostevPRA2023}:
\begin{equation}
P_{\rm true}(K,\phi)
=\eta^{-K}\!\sum_{N=K}^{N_{\max}}
(-1)^{N-K}\binom{N}{K}\left(\tfrac{1-\eta}{\eta}\right)^{N-K}
P_{\det}(N,\phi).
\
\label{eq:coef-inverse-final}
\end{equation}
Equations \eqref{eq:Pdet-forward-final},~\eqref{eq:coef-inverse-final} were used to produce Figure ~\ref{fig:fringe-loss-inverse}, where an ideal FRINGE trace of a random quantum state was synthetically thinned by a quantum efficiency $\eta=0.5$ and then recovered exactly by applying the inversion. 

We emphasize that in the presence of realistic measurement noise, not every level of loss can be reliably corrected. The inverse map in Eq.~\eqref{eq:coef-inverse-final} amplifies technical noise that is present, to some extent, in any measurement. Therefore, in practice, one should determine empirically which loss levels can be consistently corrected. A natural consistency check is to first apply the inverse map to obtain a loss-corrected trace, then reapply the loss transformation of Eq.~\eqref{eq:Pdet-forward-final} and verify that this round trip reproduces the measured dataset within the expected noise level. Only when this test is satisfied should the loss-corrected trace be used as the basis for quantum-state reconstruction.

\textbf{Mixed states. } For a mixed state $\rho$, the FRINGE trace is given by
\begin{equation}
\label{eq:balanced-fringe}
\begin{aligned}
P(N,\phi)
&= 2^{-N}\!\!\sum_{m,m'=0}^{N}\!\sqrt{\binom{N}{m}\binom{N}{m'}}\,
e^{i(m-m')\phi}\,\rho_{m m'}\\
&\quad\times\,\rho_{N-m,\,N-m'} .
\end{aligned}
\end{equation}

To implement the reconstruction of $\rho$ from $P(N,\phi)$ numerically, we represent a ground-truth density matrix $\rho_{\rm true}$ on a truncated $d$-dimensional Fock Hilbert space and compute a synthetic FRINGE trace via Eq.~\eqref{eq:balanced-fringe}. To recover the populations $\rho_{nn}$, we average the FRINGE trace over $\phi$ to obtain the DC harmonic $h_N=\langle P(N,\phi)\rangle_\phi$,  
\begin{equation}
h_N = 2^{-N}\sum_{m=0}^{N}\binom{N}{m}\,\rho_{N-m,N-m}\,\rho_{m,m}.
\end{equation}
This gives a recursion for the density-matrix diagonal:
\begin{equation}
\rho_{00}=\sqrt{h_0},\qquad
\rho_{nn}=\frac{2^{\,n}h_n-\sum_{m=1}^{n-1}\binom{n}{m}\rho_{n-m,n-m}\rho_{mm}}
{2\,\rho_{00}} .
\end{equation}

The coherences $\rho_{nm}$ are obtained through numerical optimization. To keep the optimization unconstrained while guaranteeing physicality, we parameterize the state by a complex matrix $T\in\mathbb C^{d\times d}$ and set
\cite{JamesPRA2001}
\begin{equation}
\rho(T)=\frac{T\,T^\dagger}{\operatorname{Tr}(T T^\dagger)}.
\end{equation}
This is a Cholesky-like parameterization: for any choice of $T$, the matrix $\rho(T)$ is automatically positive semidefinite and normalized to unit trace.  Thus the least-squares fit can be performed over unconstrained complex parameters in $T$ while ensuring that the reconstructed $\rho$ is a valid density matrix. Given $T$, we compute $\rho(T)$ and subsequently $P(N,\phi)$, denoting this prediction by $P_{\rm pred}(T)$. We minimize the sum of squared residuals over all $N$ and $\phi$, plus a small quadratic penalty that keeps the fitted diagonal near the analytically reconstructed $\rho_{nn}$: 
\begin{equation}
\label{eq:objective}
\begin{aligned}
\mathcal J(T)
&= \sum_{N,\phi}\!\left|P_{\rm meas}(N,\phi)-P_{\rm pred}(N,\phi;T)\right|^2\\
&\quad+\,\lambda_{\rm diag}\sum_{n=0}^{d-1}\left|\rho_{nn}(T)-\rho_{nn}\right|^2 ,
\end{aligned}
\end{equation}
with $\lambda_{\rm diag}=0.01$. We minimize the cost function $\mathcal J(T)$ using a nonlinear least-squares optimizer. 

\begin{figure}[t]
  \centering
  \includegraphics[width=\linewidth]{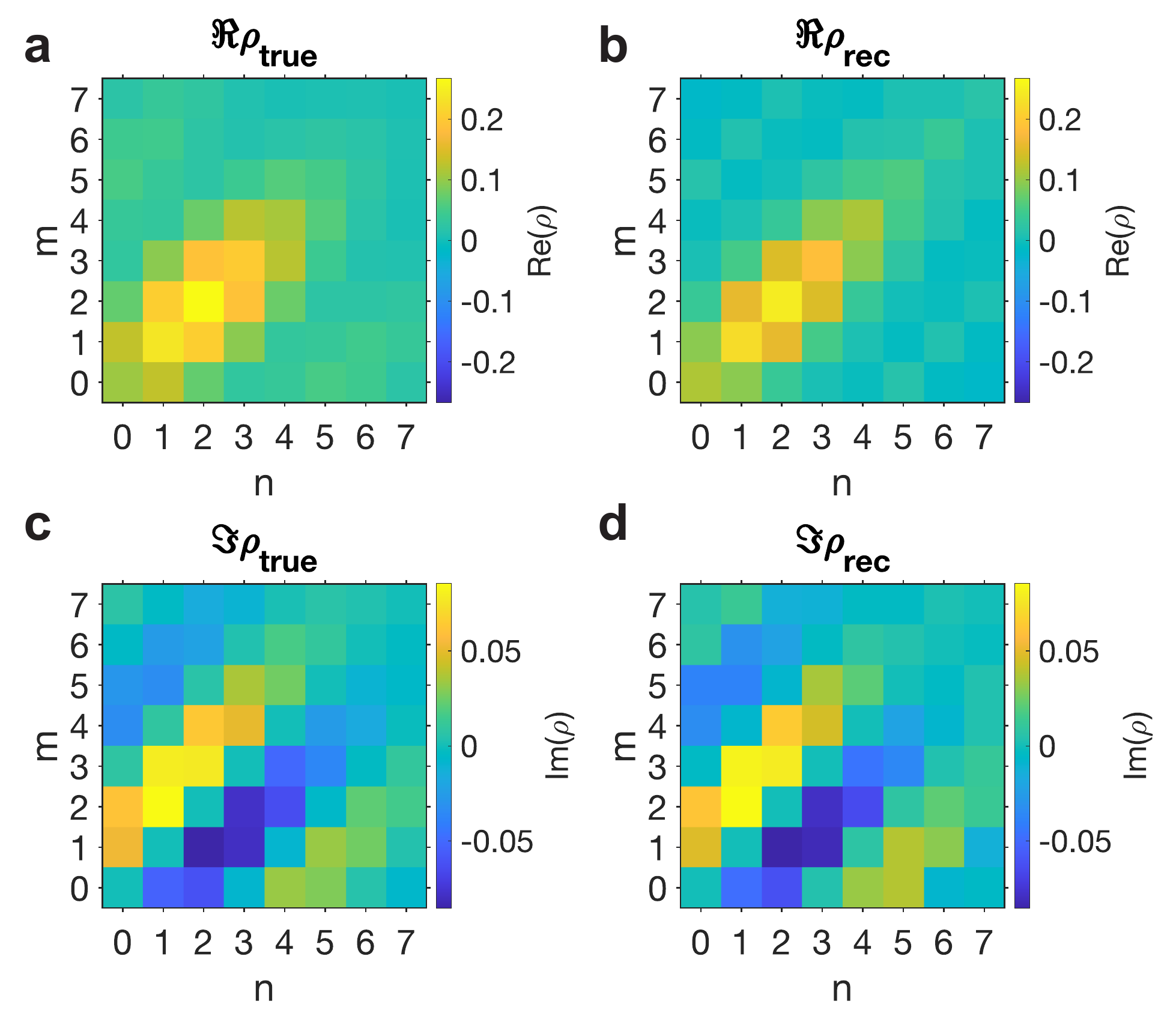}
  \caption{\textbf{FRINGE reconstruction of a mixed state.}
  (a)–(b) Real parts of the true and reconstructed density matrices $\Re(\rho_{\rm true})$ and $\Re(\rho_{\rm rec})$; (c)–(d) imaginary parts $\Im(\rho_{\rm true})$ and $\Im(\rho_{\rm rec})$. 
  Data were simulated for a mixture of two coherent states in a truncated Fock space ($d=8$). The FRINGE trace $P(N,\phi)$ was generated from Equations \eqref{eq:balanced-fringe}. The density matrix was then recovered by nonlinear least squares over the parameterization $\rho=T T^\dagger/Tr(TT^\dagger)$, with a small diagonal anchor extracted from the trace.}
  \label{fig:loss-div-fringe}
\end{figure}

As a concrete example, we apply this procedure to a mixed state composed of two coherent states,  
\(\rho_{\rm true}=p\,\ket{\alpha}\!\bra{\alpha}+(1-p)\,\ket{\beta}\!\bra{\beta}\) (normalized such that \(\operatorname{Tr}\rho_{\rm true}=1\)).  
In the numerical demonstration we consider the Hilbert space dimension \(d=8\), and coherent state parameters \(\alpha=1.5\), \(\beta=1.5\,e^{i\,1.0}\), with \(p=0.6\).  
Figure~\ref{fig:loss-div-fringe} compares the ground-truth density matrix with the reconstructed density matrix obtained by minimizing Eq.~\eqref{eq:objective}; the close agreement demonstrates that FRINGE provides sufficient information to accurately recover both populations and coherences of mixed states.

We note that the interference of identical copies of a quantum state was shown in the past to enable estimation of nonlinear functionals of the state, such as purity and non-classicality witnesses of the density matrix~\cite{EkertPRL2002_DirectFunctionals, GriffetPRA2023_QCS,ArnhemPRA2022_MulticopyNonclassical}. The FRINGE protocol presented above is conceptually close in spirit to these multicopy and interferometric approaches: it also uses two identical copies of a quantum state and photon-number-resolved detection. However, the goal is different. Refs.~\cite{EkertPRL2002_DirectFunctionals,GriffetPRA2023_QCS,ArnhemPRA2022_MulticopyNonclassical} focus on extracting specific nonlinear functionals of the density, and Ref.~\cite{SahooPRL2020_QStateInterferography} targets qubit and qudit states, focusing on the polarization degree of freedom. By contrast, FRINGE uses a Young double-slit geometry and the  two-dimensional dataset $P(N,\phi)$ to perform  quantum state tomography of an arbitrary single-mode quantum state in the Fock basis.
Experimentally, FRINGE and the approaches of Refs.~\cite{EkertPRL2002_DirectFunctionals,GriffetPRA2023_QCS,ArnhemPRA2022_MulticopyNonclassical} can be implemented in the same setup, enabling direct cross-checks and consistency tests between the methods.
To conclude, we have shown that a photon–number–resolved Young interferometer provides an informationally complete, self–referenced measurement for quantum states of paraxial light beams. The measurement acts as a projection on the states $\Pi^{(N)}(\phi)=\ket{N;\phi}\!\bra{N;\phi}$, and the resulting two–dimensional data $P(N,\phi)$ are mathematically analogous to a FROG trace. From the $\phi$–harmonics of $P(N,\phi)$ we derived a procedure for the reconstruction of the photon number probability amplitudes $c_n$. The reconstruction is unique up to the unavoidable ambiguities of self–referenced interferometry: a global/linear phase ramp $c_n\mapsto e^{i(\chi+n\theta)}c_n$ and global conjugation $c_n\mapsto c_n^\ast$. Numerical tests on a squeezed–coherent state confirm quantitative agreement between the recovered $\{c_n\}$ and the ground truth and, crucially, between the true and predicted distributions $P(N,\phi)$. For mixed states, we demonstrate numerically that a single balanced FRINGE trace provides sufficient information for state reconstruction, recovering both populations and coherences.
The ability to reconstruct full Fock--basis wavefunctions and density matrices from a single self-interferometric dataset offers a versatile route to characterization of multi-photon quantum light sources across spectral, temporal, and spatial regimes. By leveraging its direct analogy to FROG, FRINGE can import decades of algorithmic developments from ultrafast optics---including ptychography, vectorial extensions, and machine-learning--assisted retrieval schemes, into quantum optics \cite{SeifertJOSAB2004,KanePCGPA2008,SidorenkoOptica2016,genty2021machine, VFROG}. FRINGE provides quantum-state estimation that is reference-free, intrinsically mode-matched, and applicable across any spectral range where photon-counting detectors are available.

\begin{acknowledgments}
M.E.T. is grateful to Carlos Trallero-Herrero for inspiring and insightful discussions. 

\end{acknowledgments}

\balance

\bibliographystyle{apsrev4-2}
\bibliography{References}

@article{Fock_LvovskyPRL2001,
  title={Quantum state reconstruction of the single-photon Fock state},
  author={Lvovsky, Alexander I and Hansen, Hauke and Aichele, T and Benson, O and Mlynek, J and Schiller, S},
  journal={Physical Review Letters},
  volume={87},
  number={5},
  pages={050402},
  year={2001},
  publisher={APS}
}

@article{DFock_LvovskyBabichevPRA2002,
  title={Synthesis and tomographic characterization of the displaced Fock state of light},
  author={Lvovsky, AI and Babichev, SA},
  journal={Physical Review A},
  volume={66},
  number={1},
  pages={011801},
  year={2002},
  publisher={APS}
}

@article{Squeezed_BreitenbachNature1997,
  title={Measurement of the quantum states of squeezed light},
  author={Breitenbach, Gerd and Schiller, Stephan and Mlynek, J{\"u}rgen},
  journal={Nature},
  volume={387},
  number={6632},
  pages={471--475},
  year={1997},
  publisher={Nature Publishing Group UK London}
}

@article{LowEff_EspositoNJP2014,
  title={Pulsed homodyne Gaussian quantum tomography with low detection efficiency},
  author={Esposito, Martina and Benatti, Fabio and Floreanini, Roberto and Olivares, Stefano and Randi, Francesco and Titimbo, Kelvin and Pividori, Marco and Novelli, Fabio and Cilento, Federico and Parmigiani, Fulvio and others},
  journal={New Journal of Physics},
  volume={16},
  number={4},
  pages={043004},
  year={2014},
  publisher={IOP Publishing}
}

@article{LvovskyRaymerRMP2009,
  title={Continuous-variable optical quantum-state tomography},
  author={Lvovsky, Alexander I and Raymer, Michael G},
  journal={Reviews of modern physics},
  volume={81},
  number={1},
  pages={299--332},
  year={2009},
  publisher={APS}
}

@book{Leonhardt1997,
  title={Measuring the quantum state of light},
  author={Leonhardt, Ulf},
  volume={22},
  year={1997},
  publisher={Cambridge university press}
}

@article{SmitheyPRL1993,
  title={Measurement of the Wigner distribution and the density matrix of a light mode using optical homodyne tomography: Application to squeezed states and the vacuum},
  author={Smithey, Daniel T and Beck, Mark and Raymer, Michael G and Faridani, Adel},
  journal={Physical review letters},
  volume={70},
  number={9},
  pages={1244},
  year={1993},
  publisher={APS}
}

@article{RiekScience2015,
  title={Direct sampling of electric-field vacuum fluctuations},
  author={Riek, Claudius and Seletskiy, Denis V and Moskalenko, Andrey S and Schmidt, JF and Krauspe, Philipp and Eckart, Sebastian and Eggert, Stefan and Burkard, Guido and Leitenstorfer, Alfred},
  journal={Science},
  volume={350},
  number={6259},
  pages={420--423},
  year={2015},
  publisher={American Association for the Advancement of Science}
}

@article{riek2017subcycle,
  title={Subcycle quantum electrodynamics},
  author={Riek, Claudius and Sulzer, Philipp and Seeger, Maximilian and Moskalenko, Andrey S and Burkard, Guido and Seletskiy, Denis V and Leitenstorfer, Alfred},
  journal={Nature},
  volume={541},
  number={7637},
  pages={376--379},
  year={2017},
  publisher={Nature Publishing Group UK London}
}

@article{hubenschmid2024optical,
  title={Optical time-domain quantum state tomography on a subcycle scale},
  author={Hubenschmid, Emanuel and Guedes, Thiago LM and Burkard, Guido},
  journal={Physical Review X},
  volume={14},
  number={4},
  pages={041032},
  year={2024},
  publisher={APS}
}

@book{GoodmanFO3e,
  title={Introduction to Fourier optics},
  author={Goodman, Joseph W},
  year={2005},
  publisher={Roberts and Company publishers}
}

@book{TrebinoBook2002,
  author = {R. Trebino},
  title = {Frequency-Resolved Optical Gating: The Measurement of Ultrashort Laser Pulses},
  publisher = {Springer},
  year = {2002},
  doi = {10.1007/978-1-4615-1181-6}
}

@article{TrebinoJOSAA1993,
  title={Using phase retrieval to measure the intensity and phase of ultrashort pulses: frequency-resolved optical gating},
  author={Trebino, Rick and Kane, Daniel J},
  journal={Journal of the Optical society of America A},
  volume={10},
  number={5},
  pages={1101--1111},
  year={1993},
  publisher={Optical Society of America}
}

@article{SeifertJOSAB2004,
  author = {B. Seifert and H. Stolz and M. Tasche},
  title = {Nontrivial ambiguities for blind frequency-resolved optical gating and the problem of uniqueness},
  journal = {J. Opt. Soc. Am. B},
  year = {2004},
  volume = {21},
  pages = {1089--1097},
  doi = {10.1364/JOSAB.21.001089}
}

@article{KanePCGPA2008,
  title={Principal components generalized projections: a review},
  author={Kane, Daniel J},
  journal={Journal of the Optical Society of America B},
  volume={25},
  number={6},
  pages={A120--A132},
  year={2008},
  publisher={OSA}
}

@article{SidorenkoOptica2016,
  title={Ptychographic reconstruction algorithm for frequency-resolved optical gating: super-resolution and supreme robustness},
  author={Sidorenko, Pavel and Lahav, Oren and Avnat, Zohar and Cohen, Oren},
  journal={Optica},
  volume={3},
  number={12},
  pages={1320--1330},
  year={2016},
  publisher={Optical Society of America}
}

@article{bendory2017uniqueness,
  title={On the uniqueness of FROG methods},
  author={Bendory, Tamir and Sidorenko, Pavel and Eldar, Yonina C},
  journal={IEEE Signal Processing Letters},
  volume={24},
  number={5},
  pages={722--726},
  year={2017},
  publisher={IEEE}
}

@article{LeePRA1993,
  title={External photodetection of cavity radiation},
  author={Lee, Ching Tsung},
  journal={Physical Review A},
  volume={48},
  number={3},
  pages={2285},
  year={1993},
  publisher={APS}
}

@article{KissHerzogLeonhardtPRA1995,
  title={Compensation of losses in photodetection and in quantum-state measurements},
  author={Kiss, T and Herzog, U and Leonhardt, Ulf},
  journal={Physical Review A},
  volume={52},
  number={3},
  pages={2433},
  year={1995},
  publisher={APS}
}

@article{HerzogPRA1996,
  title={Loss-error compensation in quantum-state measurements and the solution of the time-reversed damping equation},
  author={Herzog, Ulrike},
  journal={Physical Review A},
  volume={53},
  number={3},
  pages={1245},
  year={1996},
  publisher={APS}
}

@article{GostevPRA2023,
  title={Inverse problem of photocount statistics: Applicability criterion for the inverse Bernoulli transform method},
  author={Gostev, Pavel P and Magnitskiy, Sergey A and Chirkin, Anatoly S},
  journal={Physical Review A},
  volume={107},
  number={4},
  pages={043710},
  year={2023},
  publisher={APS}
}

@article{JamesPRA2001,
  title={Measurement of qubits},
  author={James, Daniel FV and Kwiat, Paul G and Munro, William J and White, Andrew G},
  journal={Physical Review A},
  volume={64},
  number={5},
  pages={052312},
  year={2001},
  publisher={APS}
}

@article{VFROG,
  title={V-FROG—single-scan vectorial FROG},
  author={Haham, Gil Ilan and Levin, Alexander and Sidorenko, Pavel and Lerner, Gavriel and Cohen, Oren},
  journal={Journal of Physics: Photonics},
  volume={3},
  number={3},
  pages={034017},
  year={2021},
  publisher={IOP Publishing}
}

@article{genty2021machine,
  title={Machine learning and applications in ultrafast photonics},
  author={Genty, Go{\"e}ry and Salmela, Lauri and Dudley, John M and Brunner, Daniel and Kokhanovskiy, Alexey and Kobtsev, Sergei and Turitsyn, Sergei K},
  journal={Nature Photonics},
  volume={15},
  number={2},
  pages={91--101},
  year={2021},
  publisher={Nature Publishing Group UK London}
}

@article{EkertPRL2002_DirectFunctionals,
  author    = {Artur K. Ekert and Carolina Moura Alves and Daniel K. L. Oi
               and Michal Horodecki and Pawel Horodecki and L. C. Kwek},
  title     = {Direct Estimations of Linear and Nonlinear Functionals of a Quantum State},
  journal   = {Phys. Rev. Lett.},
  volume    = {88},
  pages     = {217901},
  year      = {2002}
}

@article{GriffetPRA2023_QCS,
  author    = {Celia Griffet and Matthieu Arnhem and Stephan De Bievre and Nicolas J. Cerf},
  title     = {Interferometric Measurement of the Quadrature Coherence Scale Using Two Replicas of a Quantum Optical State},
  journal   = {Phys. Rev. A},
  volume    = {108},
  pages     = {023730},
  year      = {2023}
}

@article{ArnhemPRA2022_MulticopyNonclassical,
  author    = {Matthieu Arnhem and Celia Griffet and Nicolas J. Cerf},
  title     = {Multicopy Observables for the Detection of Optically Nonclassical States},
  journal   = {Phys. Rev. A},
  volume    = {106},
  pages     = {043705},
  year      = {2022}
}

@article{SahooPRL2020_QStateInterferography,
  author    = {R. S. Sarthour and A. C. S. de Paula and W. T. Araujo and
               D. O. Soares-Pinto and I. S. Oliveira and others},
  title     = {Quantum State Interferography},
  journal   = {Phys. Rev. Lett.},
  volume    = {125},
  pages     = {123601},
  year      = {2020}
}

@inproceedings{harrison2023generation,
  title={Generation and control of non-local quantum equivalent extreme ultraviolet photons},
  author={Harrison, Geoffrey R and Saule, Tobias and Goetz, R Esteban and Gibson, George N and Le, Anh-Thu and Trallero-Herrero, Carlos A},
  booktitle={Frontiers in Optics},
  pages={FTu6D--4},
  year={2023},
  organization={Optica Publishing Group}
}

@inproceedings{trallero2025squeezed,
  title={Squeezed and equivalent XUV photons},
  author={Trallero, Carlos A and Harrison, Geoffrey R and Gibson, George N and Le, Anh-Thu and Goetz, Esteban and Saule, Tobias},
  booktitle={DAMOP 2025},
  year={2025},
  organization={APS}
}

@article{harrison2022increased,
  title={Increased phase precision of spatial light modulators using irrational slopes: application<? TeX$\backslash$break?> to attosecond metrology},
  author={Harrison, Geoffrey R and Saule, Tobias and Davis, Brandin and Trallero-Herrero, Carlos A},
  journal={Applied Optics},
  volume={61},
  number={30},
  pages={8873--8879},
  year={2022},
  publisher={Optica Publishing Group}
}

\end{document}